\documentclass[12pt]{article}
\usepackage{graphicx,amsmath}
\newcommand{\no}{\nonumber}

\begin{document}
\begin{titlepage}
\begin{flushright}
\begin{tabular}{l}
{KEK-TH-1008}\\
{KOBE-FHD-05-01}\\
{HUPD-0501}\\
\\
\end{tabular}
\end{flushright}
\baselineskip 24pt

\begin{center}
\vspace*{0.5cm}
{\LARGE  The Decay of Tau Leptons \\ 
       Produced in Neutrino-Nucleon Scatterings}\\
\vspace*{1cm}
\baselineskip 18pt
\renewcommand{\thefootnote}{\fnsymbol{footnote}}
\setcounter{footnote}{0}
\begin{Large}
{Mayumi Aoki$^{1}$\footnote{{E-mail address}:
           \texttt{mayumi.aoki@kek.jp}}\,,
\,\,\,    Kaoru Hagiwara$^{1,2}$\,,}
\\
{Kentarou Mawatari$^{3}$\footnote{
  {E-mail address}: \texttt{kentarou@post.kek.jp}\\
         \hspace*{0.52cm} 
  {Present address}: Theory Group, KEK, Tsukuba 305-0801, JAPAN}\,,}
\,
and \,{Hiroshi Yokoya$^{4,5}$\footnote{
  {E-mail address}: \texttt{yokoya@nt.sc.niigata-u.ac.jp}\\
          \hspace*{0.52cm} 
  {Present address}: Department of Physics, Niigata University,
 Niigata 950-2181, JAPAN}}\\
\end{Large}

\vspace*{0.75cm}
{\normalsize \it $^{1}$Theory Group, KEK, Tsukuba 305-0801, JAPAN}\\
{\normalsize \it $^{2}$Department of Particle and Nuclear Physics,}\\
{\normalsize \it Graduate University for Advanced Studies, 
                 Tsukuba 305-0801, JAPAN}\\
{\normalsize \it $^{3}$Graduate School of Science and Technology,
Kobe University,}\\
{\normalsize \it Nada, Kobe 657-8501, JAPAN}\\
{\normalsize \it $^{4}$Department of Physics, Hiroshima University,}\\
{\normalsize \it Higashi-Hiroshima 739-8526, JAPAN}\\
{\normalsize \it $^{5}$Radiation Laboratory, RIKEN, Wako 351-0198, JAPAN}

\end{center}
\date{}

\vspace*{1cm}
\begin{abstract}
Energy and angular distributions of the $\tau$ decay products in the 
CERN-to-Gran Sasso $\nu_\tau$ appearance experiments are studied for 
the decay modes $\tau\to\pi\nu$ and $\tau\to\ell\bar\nu\nu$ 
($\ell=e$ or $\mu$).  
We find that the decay particle distributions in the laboratory frame 
are significantly affected by the $\tau$ polarization. 
Rather strong azimuthal asymmetry of $\pi^-$ and $\ell^-$ about the
 $\tau$ momentum axis is predicted, which may have observable
 consequences even at small statistics experiments. 
\end{abstract}
                                                                               
\thispagestyle{empty}
\end{titlepage}
\setcounter{page}{1}
\setcounter{footnote}{0}                                                      
 
\section{Introduction}

Neutrino oscillation physics is one of the most attractive field of
current particle physics, and plenty of theoretical and
experimental studies are revealing the amazing nature of the neutrino
sector, such as their non-zero masses and the large mixings. 
We are now entering the stage of precise determination of the
mass-squared differences including their signs and the mixing angles.  
The $CP$ phase of the MNS (Maki-Nakagawa-Sakata) lepton-flavor-mixing 
matrix \cite{mns} also interests us.

On the other hand, it is also important to find the direct evidence
of the neutrino oscillation within the three
generations. It can be achieved by detecting $\nu_\tau$ appearance in 
the long baseline neutrino oscillation experiments with the initial
$\nu_\mu$ beam. 
The CNGS (CERN Neutrino to Gran Sasso) long baseline experiments
\cite{cngs} with ICARUS \cite{icarus} and OPERA \cite{opera} detectors
aim to establish $\nu_\tau$ appearance by measuring the $\tau$ lepton
production events caused by the charged current (CC) interactions. They
are now under construction and plan to start taking data from the year 2006.
It is also expected that the CNGS experiments will improve significantly 
the current upper limit on $\theta_{13}$, which is the smallest of the three
mixing angles 
in the three-flavor MNS matrix, by measuring $\nu_\mu \to \nu_e$ transition
\cite{theta13}. The $\nu_\tau$ CC events followed by the $\tau\to e$ decays
contribute as background events to the signals of the $\nu_\mu\to\nu_e$ 
events. Thus, the detailed analysis of the $\nu_\tau$ CC events is
important in the CNGS experiments.

Since the $\tau$ lepton has the large mass, 
$m_\tau = 1.78$ GeV, it immediately
decays into several particles always including a neutrino ($\nu_\tau$).
For that reason, $\tau$ production will be detected through its decay
particle distributions. On the other hand, the decay particle distributions
from $\tau$ depend critically on its spin polarization. 
It is therefore important to consider the spin polarization
of $\tau$ in addition to its production cross section. 

Detailed discussions on the spin polarization of $\tau$ produced in
neutrino-nucleon scattering can be found in the recent paper \cite{taupol}.
There, the quasi-elastic scattering (QE), the resonance production (RES)
and the deep inelastic scattering (DIS) processes were considered for
the $\tau$ 
production, and it was shown that the produced $\tau^\pm$'s have 
high degree of polarization, and their spin direction depends 
non-trivially on the energy and the scattering angle of 
$\tau^\pm$ in the laboratory frame. 

In this article, we study the decay distributions from $\tau^-$ leptons
produced via the CC interactions, especially for the CNGS experiments.
We consider the $\tau^-$ production in the 
neutrino-nucleon scattering and its subsequent decays, for the decay modes 
$\tau^-\to\pi^-\nu_\tau$ and 
$\tau^-\to\ell^-\bar\nu_\ell\nu_\tau$ ($\ell=e,\,\mu$):
\begin{align}
 \nu_\tau + N \to &\ \tau^- + X;\quad  \tau^- \to
 \begin{cases}
  \pi^- + \nu_\tau \\
  \ell^- +\bar\nu_\ell +\nu_\tau  
 \end{cases} 
\end{align}

\section{Tau production}

Let us start with the brief summary of 
the cross section and spin polarization of $\tau^-$ produced via neutrino. 
One can find more details in Ref.\ \cite{taupol}.

\subsection{Kinematics and the formalism}

We consider $\tau^-$ production by the charged current (CC) reactions
off a nucleon target:
\begin{equation}
 \nu_\tau(k) + N(p) \to \tau^-(k') + X(p'). 
\end{equation}
For the hadronic final states $X$, we  consider three subprocesses;    
the quasi-elastic scattering (QE), the $\Delta$ resonance production 
(RES) and the deep inelastic scattering (DIS) processes.
The four-momenta are parametrized in the laboratory frame as 
\begin{align}
 k &= (E_\nu,\, 0,\, 0,\, E_\nu),\no \\
 p &= (M,\, 0,\, 0,\, 0),\\
 k'&= (E_\tau,\, p_\tau\sin\theta_\tau,\, 0,\, p_\tau\cos\theta_\tau),\no
\end{align}
and the following Lorentz invariant variables are introduced
\begin{align}
 & Q^2 = -q^2 = -(k-k')^2, \\
 & W^2 =(p+q)^2.
\end{align}
Each subprocess is distinguished by the hadronic invariant mass $W$: 
$W=M$ for QE, $M+m_\pi <W<W_{\rm cut}$ for RES\@. 
$W_{\rm cut}$ is an artificial boundary,  
and we regard that DIS process occurs in the regions of $W>W_{\rm cut}$%
\footnote{More detailed studies on $\tau$ productions via the CC
reactions have recently been reported in Ref.~\cite{kuzmin}.}. 
We take $W_{\rm cut}=1.4$ GeV in this report.

The differential cross section and the spin polarization vector of
produced $\tau^-$ are obtained in the laboratory frame as \cite{taupol}  
\begin{align}
 \frac{d\sigma_{\tau}}{dE_{\tau}\,d\cos\theta_{\tau}}&=
 \frac{G_{F}^{2}\kappa^{2}}{2\pi}
 \frac{p_{\tau}}{M}\,\bigg\{
 \Big(2W_{1}+\frac{m_{\tau}^{2}}{M^{2}}\,W_{4}\Big)
 \left(E_{\tau}-p_{\tau}\cos\theta_{\tau}\right)
 +W_{2}\left(E_{\tau}+p_{\tau}\cos\theta_{\tau}\right)
 \no\\ &\hspace*{70pt}
 +\frac{W_{3}}{M}\,\Big(E_{\nu}E_{\tau}+p_{\tau}^{2}
 -(E_{\nu}+E_{\tau})p_{\tau}\cos\theta_{\tau}\Big)
 -\frac{m_{\tau}^{2}}{M}\,W_{5}\bigg\}\no\\
 &\equiv \frac{G_{F}^{2}\kappa^{2}}{2\pi}
 \frac{p_{\tau}}{M}\;F,  \label{cross}
\end{align}
and
\begin{subequations}
\begin{align}
 s_{x} &=  -\,\frac{m_{\tau}\sin\theta_{\tau}}{2}
 \bigg(2W_{1}-W_{2}+\frac{E_{\nu}}{M}\,W_{3}
 -\frac{m_{\tau}^{2}}{M^{2}}\,W_{4}+\frac{E_{\tau}}{M}\,W_{5}\bigg)
 \bigg{/}F,\\
 s_{y} &=0,\\
 s_{z} &= -\,\frac{1}{2}\bigg\{
 \Big(2W_{1}-\frac{m_{\tau}^{2}}{M^{2}}\,W_{4}\Big)
 \left(p_{\tau}-E_{\tau}\cos\theta_{\tau}\right)
 +W_{2}\left(p_{\tau}+E_{\tau}\cos\theta_{\tau}\right)
 \no\\&\hspace*{40pt}
 +\frac{W_{3}}{M}\,\Big((E_{\nu}+E_{\tau})p_{\tau}
 -(E_{\nu}E_{\tau}+p_{\tau}^{2})\cos\theta_{\tau}\Big)
 -\frac{m_{\tau}^{2}}{M}\,W_{5}\cos\theta_{\tau}\bigg\}\bigg{/}F,
\end{align}\label{spin}%
\end{subequations}
where $G_F$ is the Fermi constant and $\kappa=M_W^2/(Q^2+M_W^2)$. 
$\vec{s}=(s_x,s_y,s_z)$ is defined in the $\tau$ rest frame in which the
$z$-axis is taken along its momentum direction and the $y$-axis is along
$\vec k \times \vec{k'}$, the normal of the scattering plane, in the
laboratory frame.  
It is  normalized as $|\vec{s}|=1/2$ for pure spin eigenstates.
$W_{i=1,\ldots,5}$ are structure functions defined 
with the generic decomposition of the hadronic tensor,
\begin{multline}
 W_{\mu\nu}(p,q)
 =-g_{\mu\nu}W_1(p\cdot q,\,Q^2)
 +\frac{p_\mu p_\nu}{M^2}\,W_2 (p\cdot q,\,Q^2)
 -i\epsilon_{\mu\nu\alpha\beta}\frac{p^\alpha q^\beta}
 {2M^2}\,W_3 (p\cdot q,\,Q^2) \\
  +\frac{q_\mu q_\nu}{M^2}\,W_4 (p\cdot q,\,Q^2)
  +\frac{p_\mu q_\nu+q_\mu p_\nu}{2M^2}\,W_5 (p\cdot q,\,Q^2),
\end{multline}
where the totally anti-symmetric tensor $\epsilon_{\mu\nu\alpha\beta}$ is 
defined as $\epsilon_{0123}=1$. These functions can be estimated for
each process, QE, RES, and DIS, as follows.

\subsection{Hadronic tensor}
\subsubsection{Quasi-elastic scattering (QE)}

The hadronic tensor for the QE process,
$\nu_{\tau} + n \to \tau^- +p$,
is written by using the hadronic weak transition current
$J_\mu$ as follows \cite{llewellyn}:
\begin{equation}
 W^{\rm QE}_{\mu\nu}=\frac{\cos^2\theta_c}{4}\sum_{\rm spins}
 J_\mu{J_\nu}^*\,\delta(W^2-M^2),
\end{equation}
where $\theta_c$ is the Cabibbo angle. $J_\mu$ is defined as
\begin{equation}
 J_\mu\equiv\langle p(p')|\hat{J}_{\mu}|n(p) \rangle=
       \bar u_p(p')\,\Gamma_\mu\,u_n(p),
\end{equation}
where $\Gamma_\mu$ is written in terms of the six weak form factors 
of the nucleon, $F^V_{1,2,3}$, $F_A$, $F_3^A$ and $F_p$, as
\begin{equation}
 \Gamma_\mu =
  \gamma_\mu F^V_1 
 +\frac{i\sigma_{\mu\alpha}q^\alpha\xi}{2M}F^V_2
 +\frac{q_\mu}{M}F^V_3 + 
  \left[ \gamma_\mu F_A
 +\frac{(p+p')_\mu}{M}F^A_3
 +\frac{q_\mu}{M}F_p \right]\gamma_5.
\end{equation}
We can drop $F^V_3$ and $F^A_3$ because of the time reversal invariance 
and the isospin symmetry. Moreover, the vector form factors
$F_1^V$ and $F_2^V$ are related to the electromagnetic form factors 
of nucleons under the conserved vector current (CVC) hypothesis:
\begin{equation}
 F^V_1(q^2)= 
  {G^V_E-\frac{q^2}{4M^2}G^V_M \over 1-\frac{q^2}{4M^2}}, \quad
 \xi F^V_2(q^2)=\frac{ G^V_M-G^V_E}{ 1-\frac{q^2}{4M^2}},
\end{equation}
where 
\begin{eqnarray}
 G^V_E(q^2)=\frac{G^V_M(q^2)}{1+\xi}=\frac{1}{(1-q^2/M_V^2)^2}
\end{eqnarray}
with a vector mass $M_V=0.84$ GeV and $\xi=3.706$. 
For the axial-vector form factor $F_A$ and the pseudoscalar form factor
$F_p$, we use 
\begin{eqnarray}
 F_A(q^2)=\frac{F_A(0)}{(1-q^2/M_A^2)^2}, \quad
 F_p(q^2)=\frac{2M^2}{m_\pi^2-q^2}F_A(q^2),
\end{eqnarray}
with $F_A(0)=-1.27$ \cite{pdg} and an axial-vector mass $M_A=1.026$ 
GeV \cite{bernard}.
Notice that the pseudoscalar form factor $F_p$ plays an important 
role for the polarization of $\tau$ produced by neutrino
because its contribution is proportional to the lepton mass and 
it has the spin-flip nature, although it is not known well experimentally;
see Ref.\ \cite{psff} for details.

\subsubsection{Resonance production (RES)}

The hadronic tensor for the $\Delta$ resonance production (RES) process,
$\nu_{\tau}+n\ (p) \to \tau^-+\Delta^+\ (\Delta^{++})$,
is calculated in terms of the $N$-$\Delta$ weak transition current 
$J_{\mu}$ as follows \cite{llewellyn,schreiner,singh}:
\begin{equation}
 W^{\rm RES}_{\mu\nu}=
 \frac{\cos^2\theta_c}{4}\sum_{\rm spins}J_\mu {J_\nu}^*\, 
 \left|{\sqrt{W\Gamma (W)/\pi} \over W^2-M_{\Delta}^2+iW \Gamma (W)}
 \right|^2
\end{equation}
with the running width
\begin{equation}
 \Gamma(W) = 
  \Gamma(M_{\Delta})\,\frac{M_{\Delta}}{W}\,\frac{\lambda^{\frac{1}{2}}
 (W^2,M^2,m_{\pi}^2)}{\lambda^{\frac{1}{2}}(M_{\Delta}^2,M^2,m_{\pi}^2)}. 
\end{equation}
$\Gamma(M_{\Delta})=0.12$ GeV and $\lambda(a,b,c)=a^2+b^2+c^2-2(ab+bc+ca)$.
The current $J_{\mu}$ for the process 
$\nu_{\tau}+n\to\tau^{-}+\Delta^{+}$ is parametrized as 
\begin{equation}
 J_{\mu}\equiv\langle\Delta^{+}(p')|\hat{J}_{\mu}|n(p) \rangle 
 =\bar{\psi}^{\alpha}(p')\,\Gamma_{\mu\alpha}\,u_{n}(p),
\end{equation}
where $\psi^{\alpha}$ is the spin-3/2 particle wave function and 
the vertex $\Gamma_{\mu\alpha}$ is expressed in terms of 
the eight weak form factors 
$C^{V,A}_{i=3,4,5,6}$ as 
\begin{align}
 \Gamma_{\mu\alpha} &= \left[
 \frac{g_{\mu\alpha}\!\not\!q-\gamma_{\mu}q_{\alpha}}{M} C^V_3
 +\frac{g_{\mu\alpha}\,p'\cdot q-p'_{\mu}q_{\alpha}}{M^2} C^V_4 
 +
 \frac{g_{\mu\alpha}\, p\cdot q - p_{\mu}q_{\alpha}}{M^2} C^V_5
 +\frac{q_{\mu}q_{\alpha}}{M^2} C^V_6 \right]\gamma_5 \no\\
 &\qquad+
 \frac{g_{\mu\alpha}\!\not\!q -\gamma_{\mu}q_{\alpha}}{M} C^A_3
 +\frac{g_{\mu\alpha}\,p'\cdot q -p'_{\mu}q_{\alpha}}{M^2} C^A_4 
 +
  g_{\mu\alpha} C^A_5
 +\frac{q_{\mu}q_{\alpha}}{M^2}  C^A_6.
\end{align}
By using the isospin invariance and the Wigner-Eckart theorem, 
we obtain another $N$-$\Delta$ weak transition current as 
$ \langle\Delta^{++}|\hat{J}_{\mu}|p \rangle = 
 \sqrt{3}\langle\Delta^{+}|\hat{J}_{\mu}|n \rangle$.
From the CVC hypothesis, $C^{V}_{6}=0$ and the other vector 
form factors $C^{V}_{i=3,4,5}$ are related to 
the electromagnetic form factors. 
Assuming the magnetic dipole dominance \cite{ruso}, 
we have $C^V_5=0$ and $C^V_4=-\frac{M}{M_\Delta}\,C^V_3$.
For $C_3^V$, we adopt the modified dipole 
parameterizations \cite{pys,olsson}:
\begin{equation}
  C^V_3(q^2)=\frac{C_3^V(0)}{\left(1-\frac{q^2}{M_V^2}\right)^2}\,
  { 1\over 1-\frac{q^2}{4M_V^2}}
\end{equation}
with $C_3^V(0)=2.05$ and $M_V=0.735$ GeV\@. 
For the axial form factors, we use \cite{pys}
\begin{eqnarray}
 C^A_5(q^2) = {C_5^A(0) \over (1- {q^2\over M_A^2})^2} 
     {1 \over 1-{q^2 \over 3M_A^2}}, \quad
 C^A_6(q^2)=\frac{M^2}{m_{\pi}^2-q^2} C_5^A(q^2),
\end{eqnarray}
with $C_5^A(0)=1.2$ and $M_A=1.0$ GeV\@.
For $C_3^A$ and $C_4^A$, $C_3^A=0$ and $C_4^A=-{1\over 4}C_5^A$ give 
good agreements with the data \cite{schreiner}.
As in the case of $F_{p}(q^{2})$ in the QE process, the pseudoscalar
form factor $C_{6}^{A}(q^{2})$ has significant effects on the $\tau$
production cross section and the $\tau$ polarization \cite{psff}.

\subsubsection{Deep inelastic scattering (DIS)}

In the DIS region, the hadronic tensor is estimated by using the
quark-parton model:
\begin{equation}
 W^{\rm DIS}_{\mu\nu}(p,q)=\sum_{q,\bar q}\int\frac{d\xi}{\xi}
 f_{q,\bar q}(\xi,Q^2) K^{(q,\bar q)}_{\mu\nu}(p_q,q),
\end{equation}
where $p_q=\xi p$ is the four-momentum of the scattering quark, 
$\xi$ is its momentum fraction, and  
$f_{q,\bar q}$ are parton distribution functions (PDFs) 
inside a nucleon. The quark tensor $K^{(q,\bar{q})}_{\mu\nu}$ is 
\begin{align}
 K^{(q,\bar q)}_{\mu\nu} (p_q,q)=&
 \delta(2\,p_q\cdot q-Q^{2}-m_{q'}^2)\no \\
 &\times 2[-g_{\mu\nu}(p_q\cdot q)+2p_{q\mu}p_{q\nu} 
 + p_{q\mu}q_\nu +q_\mu p_{q\nu}
 \mp i\epsilon_{\mu\nu\alpha\beta}p_{q}^\alpha q^\beta].
\end{align}
The upper sign should be taken for quarks and the lower sign for antiquarks. 
We retain the final quark mass $m_{q'}$ for the charm quark as  
$m_{c}=1.25$ GeV, but otherwise we set $m_{q'}=0$. 
In the calculation, we used MRST2002 \cite{mrst} for the PDFs%
\footnote{We adopt a naive extrapolation of the parton model calculation
for low $Q^2$ in the $W>W_{\rm cut}$ region, 
by freezing the PDFs when $Q^2<Q^2_0$ (=1.25 GeV$^2$).}.    

By neglecting both the nucleon mass and the initial quark masses,   
the following relations are obtained: 
\begin{equation}
 W_{1}(p\cdot q,Q^{2})=F_{1}(\xi,Q^{2}),
 \quad  W_{i=2,\ldots,5}(p\cdot q,Q^{2})= 
 {M^2 \over p\cdot q }\,  F_{i=2,\ldots,5}(\xi,Q^{2}).
\end{equation}
Here,
\begin{gather}
 F_{1}=\sum_{q,\bar{q}}f_{q,\bar{q}}(\xi,Q^{2}),\quad 
  F_{2}=2\sum_{q,\bar{q}}\xi\,f_{q,\bar{q}}(\xi,Q^{2}), \no\\
 F_{3}=2\sum_{q}f_{q}(\xi,Q^{2})
 -2\sum_{\bar{q}}f_{\bar{q}}(\xi,Q^{2}),\quad
  F_{4}=0,\quad
  F_{5}=2\sum_{q,\bar{q}}f_{q,\bar{q}}(\xi,Q^{2}),  
\end{gather}
where $\xi=Q^2/2p\cdot q \equiv x$ for massless final quarks  
($m_{q'}=0$), and  
$\xi = {x}/{\lambda}$ with $\lambda = Q^{2}/(Q^{2}+m_{q'}^{2})$ for $q'=c$.
In fact, the differential cross section, Eq.\ (\ref{cross}), does not satisfy 
the positivity condition near the threshold with this naive
replacement. We modify the $W_1$ structure function as
$W_1= (1+{\xi M^2 \over p\cdot q})F_1$ in order to preserve the
positivity \cite{taupol}.

\subsection{Polarization of produced $\tau^-$ at a fixed neutrino energy}

We summarize the cross section and spin polarization of $\tau^-$ 
produced in neutrino-nucleon scattering for isoscalar targets
at the incident neutrino energy $E_\nu=10$ GeV 
on the $p_\tau\cos\theta_\tau$-$p_\tau\sin\theta_\tau$ plane, 
where $p_\tau$ and $\theta_\tau$ are the $\tau$ momentum 
and the scattering angle in the laboratory frame. 

\begin{figure}[h]
\begin{center}
 \includegraphics[width=6.25cm,clip]{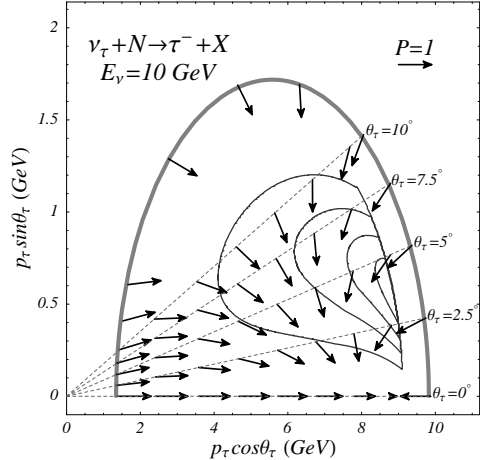}
\end{center}
\vspace*{-0.5cm}
\caption{
The contour map of the DIS cross section on the   
$p_{\tau}\cos\theta_\tau$-$p_{\tau}\sin\theta_\tau$ plane for the  
$\nu_{\tau}N\to \tau^{-}X$ process at $E_{\nu}=10$ GeV in the laboratory frame.
The kinematical boundary is shown by the thick curve. The QE process 
contributes along the boundary, and the RES process contributes just inside 
of the boundary. The $\tau$ polarization are shown by the arrows. 
The length of the arrows give the degree of
 polarization, and the direction of arrows give that of the $\tau$
spin in the $\tau$ rest frame. The size of the 100\% polarization ($P=1$) 
arrow is shown as a reference. The arrows are shown along the laboratory 
scattering angles, $\theta_\tau=0^{\circ}$, $2.5^{\circ}$, $5^{\circ}$, 
$7.5^{\circ}$, and $10^{\circ}$, as well as along the kinematical
 boundary.}\label{convec}
\end{figure}

In Fig.\ \ref{convec}, the differential cross sections
$d\sigma/dp_z\,dp_T$, obtained from Eq.\ (\ref{cross}), 
are shown as a contour map%
\footnote{It must be noted that this contour map differs from
Fig.\ 11 of Ref.\ \cite{taupol}, where
$d\sigma/dE_\tau\,d\cos\theta_\tau$ is plotted.},
where $p_z=p_\tau \cos\theta_\tau$ and $p_T=p_\tau\sin\theta_\tau$.
Only the contours of the DIS cross section are plotted to avoid too much
complexity. 
Each contour gives the value of the differential cross section in
the unit of fb/GeV$^2$; the outermost line is 1 fb/GeV$^2$ and  
the innermost line is for 4 fb/GeV$^2$.
The kinematical boundary is shown by the thick curve. The QE process 
contributes along the boundary, and the RES process contributes just
inside of the boundary. 
The contour map shows that the contributions in the forward scattering angles
in the larger $p_\tau$ side are important. In that region,
the cross sections of QE and RES are also large and comparable to that
of DIS.

The polarization vector $\vec s$, Eq.\ (\ref{spin}), of $\tau^-$ is 
also shown in Fig.\ \ref{convec}.
The length of each arrow gives the degree of 
polarization ($0\leq P=2|\vec s| \leq 1$) at each kinematical point and its 
orientation gives the spin direction in the $\tau$ rest frame. 
The produced $\tau^-$ have high degree of polarization, but their spin
directions significantly deviate from the massless limit predictions,
where all $\tau^-$ should be purely left-handed. 
Since $s_x$ of Eq.~(\ref{spin}) turns out to be always negative, the spin
vector points to the direction of the initial neutrino momentum axis. 
Qualitative
feature of the results can be understood by considering the helicity
amplitudes in the center of mass (CM) frame of the scattering particles
and the effects of Lorentz boost from the CM frame to the laboratory
frame. 
For instance, when we consider the neutrino-quark scattering in their CM frame,
produced $\tau^-$ has fully left-handed polarization at all scattering
angle due to the $V\!-\!A$ interactions and the angular momentum conservation.
After the Lorentz boost, higher energy $\tau$'s in the laboratory frame
preserve 
the left-handed polarization since those $\tau$'s have forward
scattering angles also in the CM frame. On the other hand, lower energy
$\tau$'s  in the laboratory frame tend to have right-handed
polarization since they are produced in the backward direction in the CM frame.
Also note that $s_x<0$ holds always because 
the polarization vector points to the collision point in the CM frame.
See more details in Ref.~\cite{taupol}.  
Let us stress that these features of the polarization of $\tau^-$ play
an important role in the following analysis.

\subsection{Polarization of produced $\tau^-$ in the CNGS experiments}

\begin{figure}[h]
\begin{center}
 \includegraphics[width=7cm,clip]{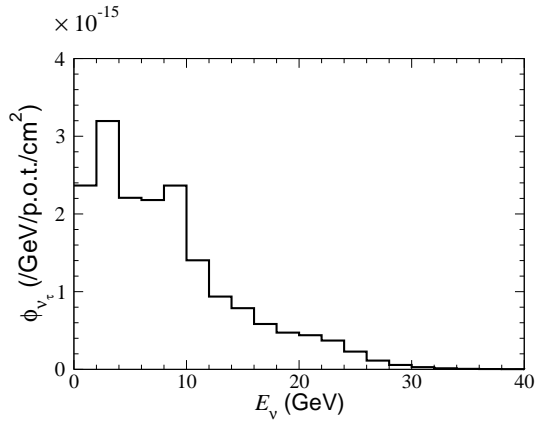}
\end{center}
\vspace*{-0.5cm}
\caption{The incoming $\nu_\tau$ flux with the
 three-neutrino model in the CNGS experiments. The neutrino oscillation
 probabilities are calculated for a set of  
parameters: $\delta m^2_{12}=8.2\times 10^{-5}\, {\rm eV}^2$ and 
$\delta m^2_{13}=2.5\times 10^{-3}\,{\rm eV}^2$,  
$\sin^2 2\theta_{12,\,23,\,13}=0.8,\,1,\,0$, respectively, the $CP$ phase
$\delta_{\rm MNS}=0^\circ$, and the matter density $\rho=3$ g/cm$^3$.}
\label{flux}
\end{figure}

\noindent
In the CNGS experiments, $\nu_\mu$ beam is produced at CERN-PS,
which is expected to deliver $4.5\times 10^{19}$ protons on target 
(p.o.t.) per year. The beam is optimized for $\nu_\tau$ appearance with
a mean neutrino energy of about 17 GeV\@.
Fig.\ \ref{flux} shows the expected $\nu_\tau$ flux
\begin{equation}
 \phi_{\nu_\tau}(E_\nu)=\sum_{\ell=e,\,\mu}\phi_{\nu_\ell}(E_\nu)
 \times P_{\nu_\ell\to\nu_\tau}(E_\nu) \label{tauflux}
\end{equation}
at Gran Sasso with the baseline length of $L=732$ km from CERN\@.
Here $\phi_{\nu_\ell}(E_\nu)$ are the initial $\nu_\ell$ fluxes 
($\ell=e,\,\mu$)
\cite{cngs} and $P_{\nu_\ell \to\nu_\tau}(E_\nu)$ are the 
$\nu_\ell \to \nu_\tau$
oscillation probabilities in the three-neutrino model.
The fraction $\nu_e/\nu_\mu$ in the initial fluxes is less than 1\%.
The neutrino oscillation probabilities are calculated for 
a set of the three neutrino model parameters:
two mass-squared differences $\delta m^2_{12,\,13}$, three mixing
angles $\theta_{12,\,23,\,13}$ and the $CP$ phase $\delta_{\rm MNS}$ 
in the MNS matrix \cite{mns}, and the matter density $\rho$.
The CHOOZ \cite{chooz} and Palo Verde \cite{paloverde}
reactor experiments give the upper bounds  
on $\sin^{2}{2\theta_{13}}$, as $\sin^{2}{2\theta_{13}}<0.16$
for $\delta m^2_{13}=2.5\times 10^{-3}\,{\rm eV^2}$.
The values of $\delta m^2_{12}$ and $\sin^{2}{2\theta_{12}}$ 
are constrained by the observations of the solar neutrinos \cite{kayser03}
and the KamLAND experiment \cite{kamland}, and these of 
$\delta m^2_{13}$ and $\sin^{2}{2\theta_{23}}$
by the atmospheric neutrino observation at Super-Kamiokande
\cite{sk-atm04} and the K2K experiment \cite{k2k}. 
No constraint on the $CP$ phase has been given by the present neutrino 
experiments.
In our analysis, we take the following values for the neutrino
oscillation parameters:
\begin{gather}
 \delta m^2_{12} = 8.2\times 10^{-5}\, {\rm eV^2},\quad
  \delta m^2_{13} = 2.5\times 10^{-3}\,{\rm eV^2},\no \\ 
 \sin^{2}{2\theta_{12}}=0.8,\quad
  \sin^{2}{2\theta_{23}}=1,\quad 
  \sin^{2}{2\theta_{13}}=0, \quad \delta_{\rm MNS}=0^\circ,
\label{osc_para}
\end{gather}
with the constant matter density of $\rho=3$ g/cm$^3$.
Here we assume the so-called normal hierarchy.
Because of setting  $\sin^{2}{2\theta_{13}}=0$, 
the neutrino oscillation probabilities are approximately described by 
those in the two neutrino ($\nu_\mu$ and $\nu_\tau$) model.
See e.g., Ref.~\cite{prosp} for details.

\begin{figure}[t]
\begin{center}
 \includegraphics[height=7cm,clip]{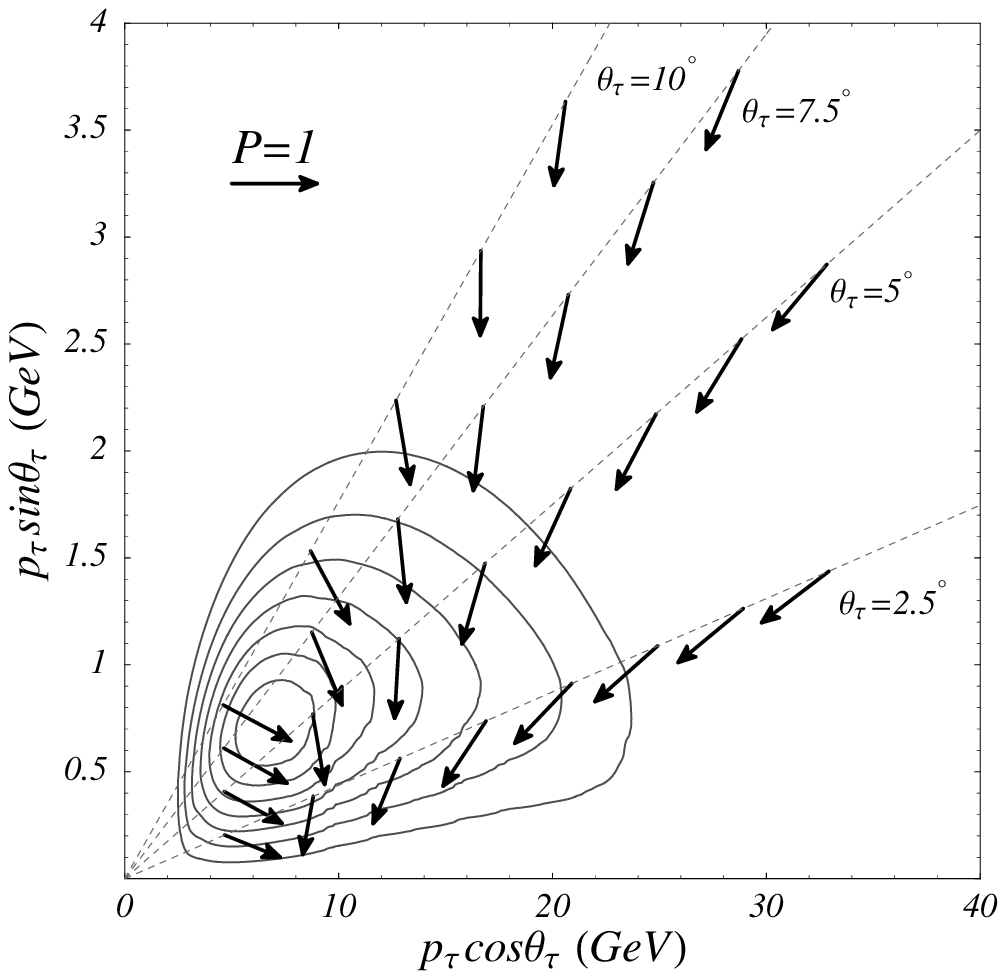}\quad 
 \includegraphics[height=7cm,clip]{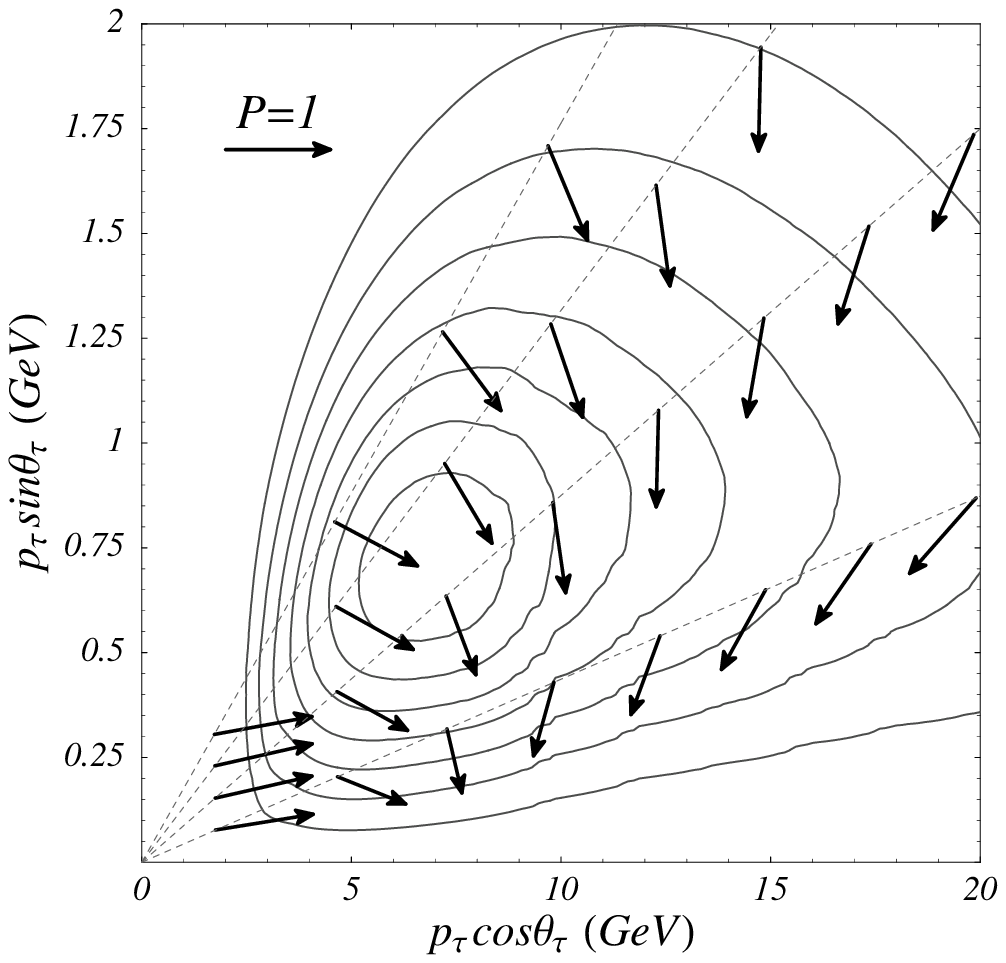}
\end{center}
\vspace*{-0.5cm}
\caption{The contour map of the number of $\tau^-$ production events on the 
$p_{\tau}\cos\theta_{\tau}$-$p_{\tau}\sin\theta_{\tau}$ plane in the CNGS
 experiments. The $\tau^-$ polarization are shown by the arrows. 
The length of the arrows give the degree of
 polarization, and the direction of arrows give that of the $\tau^-$
spin in the $\tau^-$ rest frame. The size of the 100\% polarization ($P=1$) 
arrow is shown as a reference. The arrows are shown along the laboratory 
scattering angles, $\theta_\tau=2.5^\circ$, $5^\circ$, $7.5^\circ$, and
 $10^{\circ}$. The right figure is an enlargement of the left figure.
}\label{conveccngs}
\end{figure}

Taking into account the CNGS neutrino flux shown in Fig.\ \ref{flux},
we show the distributions of events and polarization vectors of
$\tau^{-}$  on the 
$p_{\tau}\cos\theta_{\tau}$-$p_{\tau}\sin\theta_{\tau}$ plane in 
Fig.~\ref{conveccngs}. The right figure is an
enlargement of the left figure to show the polarization vectors in
detail for the important region of large cross section.
The initial neutrino energy is integrated out with the incoming $\nu_\tau$
flux, $\phi_{\nu_\tau}(E_\nu)$ of Eq.\ (\ref{tauflux}), 
whereas it is fixed at 10 GeV in Fig.\ \ref{convec}. 
The number of $\tau^-$ production events for all the QE, RES and DIS processes
are included in the contour map, where we assume 5 years 
with $4.5\times10^{19}$ p.o.t./year of the primary proton beam and
the 1.65 kton size detector, which 
are the current plan of the OPERA experiment \cite{opera}.
Each contour gives a number of events per GeV$^2$; 
the outermost line corresponds to 1 event/GeV$^2$, and  
the innermost line is for 7 events/GeV$^2$. 
The contour map shows that there are many events around $E_\tau$=10 GeV,
and around $\theta_\tau=5^\circ$. 
As for the polarization vectors, the dependence on the energy and the 
scattering angle of $\tau^-$ is rather smooth as compared to that in Fig.\
\ref{convec} because of the integration of the incident neutrino energy. 
As the right figure shows, however, the direction of the $\tau$
polarization is still non-trivial in the region which has many events.

\section{Tau decay}
\subsection{Tau decay in the $\tau$ rest frame}

Before turning to the $\tau$ decay in the CNGS experiments, we give the
formulas of the energy and angular distributions of 
the decay particles, especially $\pi^-$ and $\ell^-$, from the polarized
$\tau^-$ lepton in the $\tau$ rest frame. 
The decay distributions in the laboratory frame
can be easily obtained by simple Lorentz transformation.
Here again we note that the $z$-axis of the rest frame, 
in which we calculate the spin polarization vector of $\tau$, 
is taken along its momentum direction in the laboratory frame.

The decay distribution of $\pi^-$ via the decay mode 
$\tau^-\to\pi^-\nu_\tau$ is given as
\begin{equation}
 \frac{1}{\Gamma_\tau}\frac{d\Gamma_{\pi}}{d\hat E_{\pi}d\hat\Omega_{\pi}}=
 B_\pi\, \frac{1}{4\pi}
 \left(1+\frac{2\hat{\vec{s}}\cdot\hat{\vec{p}}_{\pi}}{\hat p_{\pi}}\right)
 \delta(\hat E_{\pi}\!-(\!m_{\tau}^2\!+\!m_{\pi}^2)/2m_{\tau}),
\end{equation}
where $\Gamma_\tau$ is the total decay width of $\tau$ and 
$B_\pi=B(\tau\to\pi\nu)=11.06\%$ \cite{pdg} is the decay branching fraction. 
All the frame dependent variables with hat ($\,\hat{}\,$) symbols
are those in the $\tau$ rest frame. 
The energy of $\pi$ is fixed as 
$\hat{E}_{\pi}=(m_{\tau}^2+m_{\pi}^2)/2m_{\tau}$, 
because of the 2-body decay kinematics. 
The angular distribution of $\pi^-$ is dictated by the $\tau^-$
polarization, and $\pi^-$ prefers to be emitted along the $\tau$
polarization direction.

Similarly, the decay distribution of $\ell^-$ via the decay mode 
$\tau^{-} \to {\ell}^{-}\bar{\nu}_{\ell}\nu_{\tau}\ (\ell=e,\mu)$ is given as
\begin{multline}
 \frac{1}{\Gamma_\tau}\frac{d\Gamma_{\ell}}{d\hat E_{\ell}d\hat\Omega_{\ell}}=
 B_\ell\,
 \frac{1}{4\pi}\frac{2}{\hat{E}^{4}_{\rm max}f(m^{2}_{\ell}/m^{2}_{\tau})}\\ 
 \times \hat p_{\ell}\hat E_{\ell}
 \left[3\hat{E}_{\rm max}-2\hat E_{\ell}-\frac{m_{\ell}^{2}}{\hat E_{\ell}}
 +\frac{2\hat{\vec{s}}\cdot\hat{\vec{p}}_{\ell}}{\hat p_{\ell}}
 \frac{\hat p_{\ell}}{\hat E_{\ell}}
 \left(\hat{E}_{\rm max}-2\hat E_{\ell}+\frac{m_{\ell}^{2}}{m_\tau}\right)\right],
\end{multline}
where $\hat{E}_{\rm max}=(m^{2}_{\tau}+m^{2}_{\ell})/2m_{\tau}$ and 
the branching fraction of $\tau\to{\ell}\bar{\nu}\nu$, $B_\ell$,  
is 17.84\% for $\ell=e$ and 17.37\% for $\ell=\mu$ \cite{pdg}.
The normalization function is $f(y)=(1-8y+8y^3-y^4-12y^2\ln{y})/(1+y)^4$. 
The distribution of $\ell^-$ depends on its energy, and has a peak in
the high energy. Furthermore, the high energy $\ell^-$ tends to be emitted
against the direction of the $\tau$ spin, although the
impact is smaller than that of the $\pi$ case. 
It is meaningful to notice that, therefore, $\pi^-$ and $\ell^-$ tend
to have the opposite preference of the $\tau$ spin dependence 
on their angular distribution.

\subsection{Tau decay in the CNGS experiments}

In this section, we present our results of the decay particle
distributions from $\tau^-$ leptons produced by the CC interactions for the
CNGS experiments. 
Main feature of our analysis is to deal with the proper spin polarization
of $\tau^-$ which is calculated for each 
production phase space, shown in Fig.\ \ref{conveccngs}.
In order to show the effects of the $\tau$ polarization on the decay
distributions, we compare the results with unpolarized $\tau$ decays and
also with completely left-handed $\tau$ decays. 

The events of the decay distributions for $\rm i = \pi,\,\ell$ are given by 
\begin{align}
 \frac{dN_{\rm i}}{dE_{\rm i}d\Omega_{\rm i}}
=& A\int_{E_{\nu}^{\rm thr}}^{E_{\nu}^{\rm max}}
\!\!dE_{\nu}\,
\phi_{\nu_{\tau}}(E_{\nu})
\int^{1}_{\rm c_{min}}\!\!d\cos\theta_{\tau}
\int_{E_{-}}^{E_{+}}\!\!dE_{\tau}
\frac{d\sigma_\tau}{dE_{\tau}d\cos\theta_{\tau}}(E_\nu)\no \\
& \times 
\frac{1}{\Gamma_{\tau}}
\frac{d\Gamma_{\rm i}}{dE_{\rm i}d\Omega_{\rm i}}
\left(E_{\tau},\theta_{\tau},
\vec{s}(E_{\nu},E_{\tau},\theta_{\tau})\right),
\label{event}
\end{align}
where $A$ is the number of active targets, $E_{\nu}^{\rm max}$ 
is the maximum value of neutrino energy in the flux, 
$E_{\nu}^{\rm thr}=m_{\tau}+m_{\tau}^2/2M$ is the threshold energy to produce 
$\tau$ lepton off a nucleon, and the other integral ranges are given by
\begin{align}
 {\rm c_{min}}(E_{\nu})&=\sqrt{1+M/E_{\nu}+M^2/E_{\nu}^2
-m_{\tau}^2/4E_{\nu}^2-M^2/m_{\tau}^2},\no\\
E_{\pm}(E_{\nu},\theta_{\tau})&=(b\pm\sqrt{b^2-ac})/a,\no
\end{align}
where $a=(E_{\nu}+M)^2-E_{\nu}^{2}\cos^{2}\theta_{\tau}$,
$b=(E_{\nu}+M)(ME_{\nu}+m_{\tau}^{2}/2)$,
$c=m_{\tau}^{2}E_{\nu}^{2}\cos^{2}\theta_{\tau}+(ME_{\nu}+m_{\tau}^{2}/2)^2$.
Here, all the frame dependent variables are defined in the laboratory
frame. $E_{\tau}$ and $\theta_{\tau}$ dependence of 
$d\Gamma_{\rm i}/dE_{\rm i}d\Omega_{\rm i}$ appear by the
Lorentz transformation from the $\tau$ rest frame to the laboratory frame. 
Although we retain only the $\pi$ or $\ell$ momenta in the above
formula, more exclusive measurements may be possible. For example, 
a well-developed emulsion detector used in OPERA experiment \cite{opera} 
can detect the kink of $\tau$ decay and can measure the 
scattering angle of $\tau$ with the accuracy of mrad, and also the
$\tau$ flight length is measurable. 
Therefore it would be possible to do more detailed analysis of $\tau$ events, 
if there were enough statistics. 

\begin{figure}[t]
\begin{center}
 \includegraphics[width=10cm,clip]{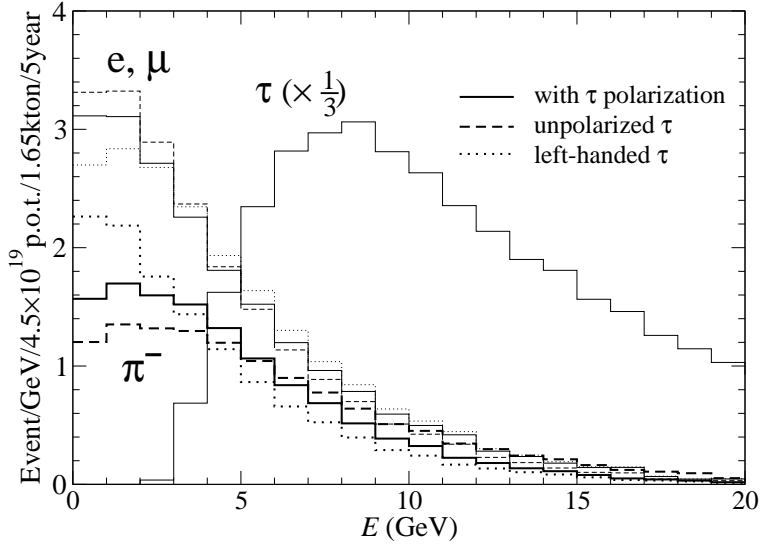}
\end{center}
\vspace*{-0.5cm}
\caption{Energy distributions of $\pi^-$ (thick lines) and $\ell^-
 (=e^-,\, \mu^-$) (medium-thick lines)
 in the decay of $\tau^-$ produced in the neutrino-nucleon CC interactions
 for the CNGS experiments. 5 years running with $4.5\times10^{19}$ p.o.t./year
 of the primary proton beam and 1.65 kton detector are assumed. 
 Solid, dashed, and dotted lines shows the energy distributions
 with the predicted $\tau$ polarization, unpolarized, 
 and purely left-handed cases, respectively. The estimated number of $\tau^-$
 production is also shown by a thin solid line with respect to the
 $\tau$ energy ($E_\tau$). 
  }\label{event1}
\end{figure}

Fig.\ \ref{event1} shows the energy distributions of $\pi^-$ (thick
lines) and $\ell^-(= e^-,\, \mu^-)$ (medium-thick lines) decayed from
$\tau^-$ produced in the neutrino-nucleon CC interactions. 
We assume the same configuration of the experimental setup as Fig.\
\ref{conveccngs}, i.e., 5 years running with $4.5\times10^{19}$ p.o.t.\
per year of the primary proton beam and 1.65 kton size detector for the
OPERA experiment \cite{opera}. (For the ICARUS experiment, 2.35 kton of
liquid Argon detector mass and 10 years running are
planned \cite{icarus}, so that more statistics are expected.)
For each decay mode, solid lines show the distributions from the decay
of $\tau^-$ with the predicted $\tau$ polarization. For comparison,  
dashed and dotted lines show those of unpolarized $\tau$
and purely left-handed $\tau$, respectively.
The estimated number of the $\tau^-$ production with respect to the
$\tau$ lepton energy, $E_\tau$, is also plotted as a thin solid line. 
The results are calculated by using Eq.\ (\ref{event}) 
with 100\% particle detection efficiency for simplicity.
For the above parameters, 113 events of $\tau^-$ are produced 
and 13 (20) of those decay into $\pi^-$ ($\ell^-$) mode. 

The $\pi^-$ and $\ell^-$ distributions have peak in the low energy
region, and in this region the polarization dependence becomes large. 
As we pointed out in the previous section, the polarization
dependence is opposite between $\pi^-$ and $\ell^-$, and is more significant 
in $\pi^-$ than in $\ell^-$. 
In the peak region, the polarization dependence affect the distribution 
around 30\% (15\%) for the $\pi^-$ ($\ell^-$) decay mode.
Expected statistics is rather small in the current design of the
CNGS experiments. However, the likelihood probability of each event will
be affected significantly by the $\tau$ polarization effects.
The characteristic feature of our prediction is
that the produced $\tau^-$ is almost fully polarized and that it has     
non-zero transverse component of the spin vector, namely $s_x$ of Eq.\
(\ref{spin}). 
The observed patterns of the $\pi^-$ and $\ell^-$ energy distributions
in the laboratory frame then follow from the energy angular
distributions in the polarized $\tau$ rest frame%
\footnote{We may note that we checked all our results by using TAUOLA (the
Monte Carlo program to simulate decays of $\tau$ leptons)
\cite{tauola}.}. 

\begin{figure}[t]
  \begin{center}
   \includegraphics[width=10cm,clip]{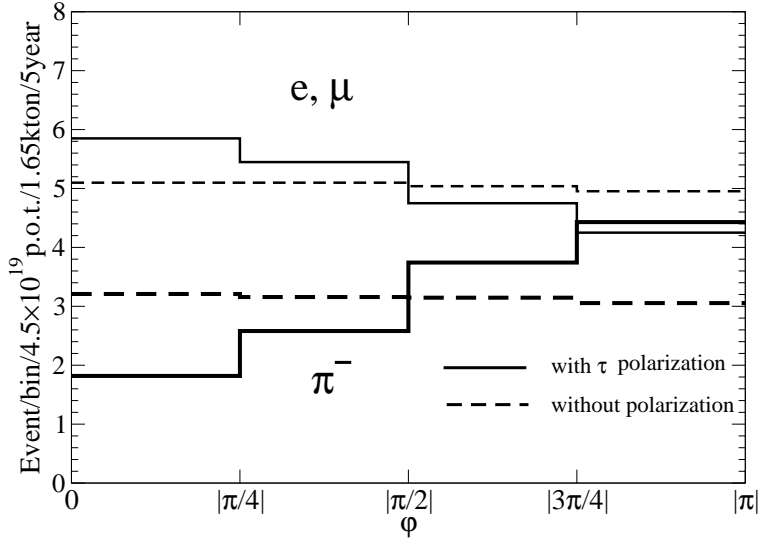}
  \end{center}
  \vspace*{-0.5cm}
  \caption{
 Azimuthal angle distribution of $\pi^-$ (thick lines) and $\ell^-$
 (thin lines). The
  setup configurations, such as neutrino flux, detector size etc., are
  the same as Fig.\ \ref{event1}. Solid lines show the distributions from
  $\tau^-$ with the predicted $\tau$ polarization, and dashed lines
  show those with the unpolarized $\tau$ case. The results of 
  purely left-handed $\tau$ are the same as those for the unpolarized
  $\tau$.
  }\label{event2}
\end{figure}

Fig.\ \ref{event2} shows the azimuthal angle distribution of $\pi^-$
(thick lines) and $\ell^-$ (thin lines). 
The azimuthal angle $\varphi_{{\rm i}=\pi,\ell}$ is given by 
$d\Omega_{\rm i}=d\cos\theta_{\rm i}\,d\varphi_{\rm i}$ in Eq.\
(\ref{event}), and is measured from the scattering plane where
$\varphi_{\rm i}=\pi/2$ is along the 
$\vec{p}_\nu \times\vec{p}_\tau$ direction in the laboratory frame,
in which the $z$-axis is taken along the direction of the $\tau$ momentum.  
Solid lines show the distributions from
$\tau^-$ with the predicted $\tau$ polarization, and dashed lines
show those from unpolarized $\tau$. The results of purely left-handed
$\tau$ are the same as those for the unpolarized $\tau$.
Since both unpolarized and purely left-handed $\tau^-$ have zero
component of perpendicular polarization, they give flat azimuthal
distributions.
The azimuthal angle distributions can be measured by tracking the
trajectory of $\tau$ leptons 
by emulsion detectors in the OPERA experiment, or by reconstructing the
hadronic cascades from neutrino-nucleon scattering. 
As is the case of energy distribution, $\pi^-$ and $\ell^-$
decay mode show the opposite feature and polarization dependence is 
clearer on $\pi^-$ mode than $\ell^-$ mode. 
At $\varphi=0$ or $|\pi|$, the dependence of the $\tau$ polarization
affects the distribution by about 47\% (16\%) for the $\pi^-$ ($\ell^-$)
decay. Even though the number of
event is limited, it may be possible to obtain a hint of such large
asymmetries. 

Finally we comment on the contributions of 
the neutrino oscillation parameters to our results.
The number of $\tau^-$ production is very sensitive to the value of 
$\delta m^2_{13}$. When we take Eq.\ (\ref{osc_para}) but 
$\delta m^2_{13} = 3\times 10^{-3}\ {\rm eV^2}$, about 50 more events
of $\tau^-$ are obtained.
On the other hand, the produced $\tau^-$ decreases (about 10\% maximum)
for the larger 
(smaller) value of $\sin^22\theta_{13}$ ($\sin^22\theta_{23}$).
However those parameters
do not change much the impacts of the $\tau^-$ polarization
on the energy and angular distributions of 
the $\tau$ decay products.

\section{Conclusion}

In this article, we have studied the effects of the spin polarization of
$\tau^-$ produced in neutrino-nucleon scattering on the subsequent decay
distributions.
The calculation of the cross section and the spin polarization of $\tau^-$ production
processes, QE, RES and DIS, were reviewed and the decay distributions of
$\tau^-$ into $\pi^-$ or $\ell^-(=e^-,\,\mu^-)$ modes were considered. 
Taking into account the polarization of produced $\tau^-$, we calculated
the energy and 
azimuthal angle distributions of $\pi^-$ and $\ell^-$ in the laboratory frame,
for the experimental setup of the CNGS long baseline project, 
OPERA and ICARUS experiments.

We found that the decay particle distributions in the laboratory frame 
are significantly affected by the $\tau^-$ polarization. 
Rather strong azimuthal asymmetry of $\pi^-$ and $\ell^-$ about the $\tau$ momentum 
axis is predicted, which may have observable consequences even at 
small statistics experiments. 

Before closing, let us mention about the decay particle distribution
from $\tau^+$, although there is no plan to use $\bar\nu_\mu$ beam in
the CNGS experiments so far. In this case the strong azimuthal asymmetry
predicted for the $\tau^-$ decay is not expected because the transverse
component of the $\tau^+$ polarization is rather small, as shown in
Fig.\ 12 of Ref.\ \cite{taupol}.

\section*{Acknowledgments}
K.M.\ would like to thank T. Morii for encouragements and S. Aoki for
valuable comments.
H.Y.\ would like to thank J. Kodaira for encouragement and 
RIKEN BNL Research Center for helpful hospitality during his stay. 
The work of K.H. is partially supported by the core university exchange
program of JSPS.


\end{document}